\documentclass[12pt]{article}
\usepackage{amsmath}
\usepackage{amssymb}
\usepackage{amsfonts}
\usepackage{epsfig}
\usepackage{mathrsfs}
\usepackage{amsmath,amsthm,amssymb,amscd}
\usepackage{MnSymbol}
\usepackage{rotating}
\usepackage{multirow}
\usepackage{float}
\usepackage[english]{babel}
\usepackage{floatpag}
\usepackage{enumerate}
\usepackage{threeparttable}
\usepackage{color}
\usepackage{tabularx}
\usepackage{booktabs}
\usepackage{lscape}
\usepackage{cite}

\newcommand{\blind}{0}

\newcommand{\ty}{\tilde{Y}}
\newcommand{\tx}{\tilde{X}}
\newcommand{\tl}{\tilde{L}}
\newcommand{\tr}{\tilde{R}}
\newcommand{\tn}{\tilde{n}}

\newcommand{\tvar}{\tilde{\varepsilon}}

\newtheorem{pro}{Proposition}
\newtheorem{thm}{Theorem}
\newtheorem{lem}{Lemma}[section]

\begin{document}

\def\spacingset#1{\renewcommand{\baselinestretch}%
{#1}\small\normalsize} \spacingset{1}


\if0\blind
{
  \title{\bf Variable Selection in Doubly Truncated Regression}
  \author{Ming Zheng, Chanjuan Lin, and Wen Yu\\
    Department of Statistics, School of Management, Fudan University}
  \maketitle
} \fi

\if1\blind
{
  \bigskip
  \bigskip
  \bigskip
  \begin{center}
    {\LARGE\bf }
\end{center}
  \medskip
} \fi

\begin{abstract}
Doubly truncated data arise in many areas such as astronomy, econometrics, and medical studies. For the regression analysis with doubly truncated response variables, the existence of double truncation may bring bias for estimation as well as affect variable selection. We propose a simultaneous estimation and variable selection procedure for the doubly truncated regression, allowing a diverging number of regression parameters. To remove the bias introduced by the double truncation, a Mann-Whitney-type loss function is used. The adaptive LASSO penalty is then added into the loss function to achieve simultaneous estimation and variable selection. An iterative algorithm is designed to optimize the resulting objective function. We establish the consistency and the asymptotic normality of the proposed estimator. The oracle property of the proposed selection procedure is also obtained. Some simulation studies are conducted to show the finite sample performance of the proposed approach. We also apply the method to analyze a real astronomical data. 
\end{abstract}

\noindent%
\vfill

Accepted by SCIENTIA SINICA Mathematica (in Chinese)

\newpage
\spacingset{1.45} 
\section{Introduction}

Truncated data arise in astronomy, econometrics, and survival analysis, etc. In truncation, only those subjects fall with an interval can be observed along with the interval. The subjects fall out of their respective intervals are not known to exist and consequently, have no chance to be observed. When the truncation interval is unbounded from above, the truncation is called left-truncation; when the interval is unbounded from below, it is called right-truncation. Double truncation occurs when the truncation interval is bounded in both sides.

Truncation may bring systematic bias to statistical analysis if it is not dealt with appropriately. For nonparametric approach of distribution estimation, Turnbull (1976) developed a general algorithm for finding the nonparametric maximum likelihood estimator for arbitrarily grouped, censored or truncated data. Lyden-Bell (1971) obtained a similar estimator for singly truncated data. The counting process techniques are then adopted by Wang et al. (1986), Keiding and Gill (1990), Lai and Ying (1991a), etc. For regression analysis with single truncation, one can refer to Bhattacharya et al. (1983) for extended Mann-Whitey approach, Tsui et al. (1988) for iterative bias adjustment, Tsai (1990) for Kendall's tau correlation, Lai and Ying (1991b) for rank-based estimation, and so on. Some more recent developments include Greene (2012) for econometrics and Kim et al. (2013) and Liu et al. (2016) for general biased sampling.

Double truncation is technically more challenging to deal with compared with single truncation, so fewer results have been found in existing literature. For distribution estimation, Efron and Petrosian (1999) and Shen (2010) developed nonparametric maximum likelihood estimation. Moreira and Alvarev (2012) proposed kernel-type density estimation approach. For two-sample problem, Bilker and Wang (1996) and Shen (2013a) extended the Mann-Whitney test. For regression ananlsis with doubly truncated responses, Moreia and Alvarev (2016) proposed kernel type approach for low dimensional covariates. Shen (2013b) considered estimating a class of semiparametric transformation models. More recently, Ying et al. (2020) proposed an extended Mann-Whitney-type loss function for parameter estimation in linear regression model with fixed covariates dimension. A Mann-Whitney-type rank estimator is then defined as its minimizer.

Besides the parameter estimation, double truncation may also bring difficulty for variable selection in regression analysis. To the best of our knowledge, no literature has studied variable selection procedure for linear regression models with doubly truncated responses. Based on the extended Mann-Whitney-type loss function proposed by Ying et al. (2020), we adopt the adaptive LASSO penalty (Zou, 2006) to develop a simultaneous estimation and variable selection procedure for the doubly truncated linear models. The proposed procedure can not only deal with fixed-dimensional covariates, but also be applicable for the diverging model size where the dimension of covariates goes to infinity with the sample size. We show that the adaptive LASSO penalty leads to the oracle properties of the selection procedure. Meanwhile, an iterative algorithm is designed for minimizing the proposed objective function. In each iteration, a least absolute deviation (LAD) optimization is involved and can be solved efficiently through standard LAD algorithm. A modified BIC approach is used for selecting the tuning parameter. We also develop a random weighting approach for estimating the standard errors of the non-zero regression parameters. Numerical studies are conducted to illustrate the finite sample behaviors of the proposed approach.

The rest of the paper is organized as follows. The next section introduces necessary notation and the estimation procedure of doubly truncated regression models. In Section 3, we give out the main results of the variable selection procedure, including the basic idea of the selection method, the algorithm, the oracle properties of the selection procedure, and the approach to determine the tuning parameter. In Section 4, the results of simulation studies are presented to show the finite sample behavior of the proposed approach and a real dataset is analyzed to illustrate the approach. Some concluding remarks are given in Section 5. All the technical details are summarized in the Appendix. 

\section{Notation, model specification, and estimation}

First some notations are introduced. Let $\ty$ be the response variable and $\tx$ be a $p$-dimensional covariate vector. We consider the linear location-shift model
\begin{eqnarray}\label{2.1}
\ty=\beta^\top\tx+\tvar,
\end{eqnarray}
where $\beta$ is the $p$-dimensional regression coefficient vector and $\tvar$ is the error term independent of $\tx$, is assumed to hold. The response $\ty$ is subject to double truncation. Let $\tl$ and $\tr$ be the left and right truncation variables, respectively. $(\ty,\tl,\tr,\tx)$ can be observed if and only if $\ty$ falls into the interval $(\tl,\tr)$. We also assume that $\ty$ is conditionally independent of $(\tl,\tr)$ given $\tx$. This is a common assumption for truncation data analysis. Under model (\ref{2.1}), this assumption is equivalent to assume that $\tvar$ is independent of $(\tl,\tr)$. The distribution and density function of $\tvar$ is denoted by $F$ and $f$, respectively.

Let $(\ty_i,\tl_i,\tr_i,\tx_i)$, $i=1,\ldots,\tn$, be independent and identically distributed (i.i.d.) copies of $(\ty,\tl,\tr,\tx)$. As we mentioned, $\ty_i$ is observed if and only if $\tl_i<\ty_i<\tr_i$, together with $(\tl_i,\tr_i,\tx_i)$. Denote the number of the observed data points by $n$, that is, $n=\#\{i:\tl_i<\ty_i<\tr_i\}$. We use $(Y_i,L_i,R_i,X_i)$, $i=1,\ldots,n$ to denote the observed data points. Moreover, for $i=1,\ldots,n$, let $e_i(\beta)=Y_i-\beta^\top X_i$, $L_i(\beta)=L_i-\beta^\top X_i$, and $R_i(\beta)=R_i-\beta^\top X_i$.

To correct the bias brought by the double truncation, Ying et al. (2020) proposed a modified Mann-Whitney-type rank estimation equation
\begin{eqnarray}\label{2.2}
\sum_{i=1}^n\sum_{j=1}^nI\{L_j(\beta)<e_i(\beta)<R_j(\beta),L_i(\beta)<e_j(\beta)<R_i(\beta)\}(X_i-X_j)\mbox{sgn}\{e_i(\beta)-e_j(\beta)\}=0,
\end{eqnarray}
where $I\{\cdot\}$ is the indicator function and $\mbox{sgn}\{\cdot\}$ is the sign function. By introducing the indicators, the estimation function takes the summation of all the so-called comparable pairs and becomes the unbiased when $\beta$ takes the true value. Meanwhile, solving (\ref{2.2}) is equivalent to minimizing the following loss function
\begin{eqnarray}\label{2.3}
l_n(\beta)=\sum_{i=1}^n\sum_{j=1}^n\left|\left[(e_i(\beta)-e_j(\beta))\wedge(R_j-Y_j)\wedge(Y_i-L_i)\right]\vee(L_j-Y_j)\vee(Y_i-R_i)\right|,
\end{eqnarray}
where $\wedge$ and $\vee$ are the minimum and maximum operator, respectively. The minimizer of $l_n(\beta)$, denoted by $\hat{\beta}$, is the proposed estimator of $\beta$ by Ying et al. (2020). They showed that the estimator is consistent and asymptotically normal under some regularity conditions.

\section{Main results of variable selection}

\subsection{The method of variable selection} 

We consider variable selection for model (\ref{2.1}) based on the observed data points. The loss function (\ref{2.3}) motivates a penalized approach for variable selection. Instead of restricting to fixed dimension of  $X_i$, we allow the dimension to go to infinity with $n$, that is, $p$ depends on $n$, denoted by $p_n$. Then all the quantities that are functions of the covariates depend on $n$. For the simplicity of notation, we shall suppress the subscript $n$ when there is no confusion. Meanwhile,  $n/\tn$ is assumed to converge to a positive constant less than 1 as $\tn$ goes to infinity. 

Because of the $L_1$ form of the loss function (\ref{2.3}), the LASSO-type penalties are naturally considered. Specifically, the objective function with usual LASSO penalty is given by 
\begin{eqnarray*}
	l_n(\beta)+\lambda_n\sum_{j=1}^{p_n}|\beta_j|,
\end{eqnarray*}
where $\lambda_n$ is a (data-dependent) tuning parameter. There exist two main concerns if we want to use the above LASSO penalty. The first one is about the computation. One can see that $l_n(\beta)$ is not convex in $\beta$, so the optimization is no longer convex. However, we note that $l_n(\beta)$ can be represented as
\begin{eqnarray*}
	l_n(\beta)=\sum_{i=1}^n\sum_{j=1}^nI\{L_j(\beta)<e_i(\beta)<R_j(\beta),L_i(\beta)<e_j(\beta)<R_i(\beta)\}|e_i(\beta)-e_j(\beta)|.
\end{eqnarray*}
The non-convexity comes from the indicator function, while $|e_i(\beta)-e_j(\beta)|$ is convex in $\beta$. Thus, one can develop an iterative algorithm where the parameter in the indicator function is kept fixed in each iteration step and the optimization is implemented for the rest part of the objective function. Some more details of the computation are discussed in Section 3.2 and Appendix A.1.

The second one is that it is well known that the LASSO estimator does not have the unbiasedness property and there are scenarios where the LASSO selection is not consistent. To overcome these drawbacks, Zou (2006) proposed adaptive LASSO penalty, in which proper weights are introduced for penalizing different coefficients in the lasso penalty. The weights used by Zou (2006) are defined based on root-$n$-consistent estimator of $\beta$. He showed that with a proper choice of the tuning parameter $\lambda_n$, the adaptive LASSO possesses the oracle properties when the dimension of $\beta$ is fixed.

Trying to keep the oracle property, we also consider the adaptive LASSO approach for variable selection. Specifically, the objective function with adaptive LASSO is given by
\begin{eqnarray}\label{3.1}
l_n^{\mbox{\tiny AL}}(\beta)=l_n(\beta)+\lambda_n\sum_{j=1}^{p_n}\hat{w}_j|\beta_j|,
\end{eqnarray}
where $\hat{w}_j$ is the adaptive weight of $\beta_j$. Originally, Zou (2006) proposed to use $\hat{w}_j=1/|\tilde{\beta}_j|^\gamma$, where $\tilde{\beta}_j$ is a root-$n$-consistent estimator of $\beta_j$ and $\gamma$ is a predetermined positive constant. In our case, apparently, even when $p_n$ diverges with $n$, we can still get $\hat{\beta}$ by minimizing $l_n(\beta)$ as long as $p_n$ does not grow too fast. Based on $\hat{\beta}$, we define $\hat{w}_j=1/|\hat{\beta}_j|^\gamma$ in (\ref{3.1}), where $\hat{\beta}_j$ stands for the $j$-th component of $\hat{\beta}$. The proposed adaptive LASSO estimator, conceptually, is defined to be the minimizer of $l_n^{\mbox{\tiny AL}}(\beta)$.

\subsection{The algorithm} 

As we mention in Section 3.1, $l_n(\beta)$ is not convex, neither is $l_n^{\mbox{\tiny AL}}(\beta)$, so we turn to an iterative procedure to get the proposed estimator. Specifically, we modify $l_n^{\mbox{\tiny AL}}(\beta)$ to
\begin{eqnarray*}
	l_n^{\mbox{\tiny AL}}(\beta,b)=\sum_{i=1}^n\sum_{j=1}^nI\{L_j(b)<e_i(b)<R_j(b),L_i(b)<e_j(b)<R_i(b)\}|e_i(\beta)-e_j(\beta)|+\lambda_n\sum_{j=1}^{p_n}\hat{w}_j|\beta_j|.
\end{eqnarray*}
Let $\hat{\beta}^{(0)}$ be an initial estimate; for example, one may use $\hat{\beta}$ as the start. Then, in the $k$-th iteration step, let $\hat{\beta}^{(k)}=\mbox{argmin}_\beta l_n^{\mbox{\tiny AL}}(\beta,\hat{\beta}^{(k-1)})$ for $k\geqslant1$. If $\hat{\beta}^{(k)}$ converges to a limit as the number of $k\to\infty$, then we use the limit as the proposed estimator. 

The main reason that we use the above iterative algorithm is because in the $k$th iteration, the objective function can be written as 
\begin{eqnarray}\label{steoLP}
l_n^{\mbox{\tiny AL}}(\beta,\hat{\beta}^{(k-1)})=\sum_{i,j\in{\cal S}_k}|e_i(\beta)-e_j(\beta)|+\lambda_n\sum_{j=1}^{p_n}\hat{w}_j|\beta_j|,
\end{eqnarray}
where ${\cal S}_k=\{(i,j) | L_j(\hat{\beta}^{(k-1)})<e_i(\hat{\beta}^{(k-1)})<R_j(\hat{\beta}^{(k-1)}),L_i(\hat{\beta}^{(k-1)})<e_j(\hat{\beta}^{(k-1)})<R_i(\hat{\beta}^{(k-1)})\}$. Thus, the optimization problem becomes a linear programming problem which can be solved efficiently. Moreover, we find that after some algebraic operations, the optimization can be transformed into a least absolute deviation (LAD) problem which can be solved efficiently through standard LAD procedure, such as the \textbf{quantreg} package in \textbf{R}. More details on the transformation are provided in the Appendix A.1.

\subsection{Large sample properties}

Next we establish the oracle property of the proposed adaptive lasso approach under some regularity conditions. The oracle property requires the consistency of $\hat{\beta}$. Although Ying et al. (2020) proved root-$n$-consistency of $\hat{\beta}$ under suitable conditions, their result is only confined to fixed dimension. When $p_n$ diverges with $n$, the consistency of $\hat{\beta}$ needs further investigation. Let $\beta_0$ denote the true value of $\beta$. The following proposition, proved in the Appendix A.2, gives out the consistency of $\hat{\beta}$ with the diverging dimension.

\begin{pro}\label{pro1}
	Assume that conditions {\it C1} to {\it C6} listed in the Appendix A.2 hold. If $p_n\log n/\sqrt{n}=o(1)$, then $\|\hat{\beta}-\beta_0\|=O_p((p_n/n)^{1/2})$.
\end{pro}

We then show the existence of an adaptive lasso estimator and it is consistent with the convergence rate of $O_p((p_n/n)^{1/2})$. The following theorem, proved in the Appendix A.2, states the results.

\begin{thm}\label{thm1}
	Assume that the conditions {\it C1} to {\it C6} listed in the Appendix A.2 hold. If $\sqrt{n}\lambda_n=o(1)$ and $p_n\log n/\sqrt{n}=o(1)$, then with probability tending to 1 there exists a local minimizer of $l_n^{\mbox{\tiny AL}}(\beta)$, denoted by $\hat{\beta}^{\mbox{\tiny AL}}$, satisfying that $\|\hat{\beta}^{\mbox{\tiny AL}}-\beta_0\|=O_p((p_n/n)^{1/2})$.
\end{thm}

To present the following oracle properties, we need some more notation. Let $\xi_{ij}(\beta)=I\{L_j(\beta)<e_i(\beta)<R_j(\beta),L_i(\beta)<e_j(\beta)<R_i(\beta)\}(X_i-X_j)\mbox{sgn}\{e_i(\beta)-e_j(\beta)\}$, $h_{ij}(\beta)=\left|\left[(e_i(\beta)-e_j(\beta))\wedge(R_j-Y_j)\wedge(Y_i-L_i)\right]\vee(L_j-Y_j)\vee(Y_i-R_i)\right|$, and $\bar{l}(\beta)=\mathsf{E}(h_{ij}(\beta))$, where the expectation is taken with respect to the distribution of the observed data. We define two $p_n\times p_n$ matrices $V=\mathsf{E}(\xi_{ij}(\beta_0)\xi_{ik}^\top(\beta_0))$ and $A=\partial^2\bar{l}(\beta)/\partial\beta\partial\beta^\top|_{\beta=\beta_0}$. We further partition $\beta_0$ into $\beta_{0,1}$ and $\beta_{0,2}$, where $\beta_{0,1}$ stands for the nonzero component of $\beta_0$ and $\beta_{0,2}$ for the zero component. The dimension of $\beta_{0,1}$ is denoted by $p_{n,1}$. Let $\hat{\beta}^{\mbox{\tiny AL}}_1$ and $\hat{\beta}^{\mbox{\tiny AL}}_2$ be the adaptive lasso estimator for $\beta_{0,1}$ and $\beta_{0,2}$, respectively. Moreover, let $V_{11}$ denote the $p_{n,1}\times p_{n,1}$ components of $V$ corresponding to $\beta_{0,1}$ and $A_{11}$ denote the $p_{n,1}\times p_{n,1}$ components of $A$.

\begin{thm}\label{thm2}
	Assume that the conditions {\it C1} to {\it C6} listed in the Appendix A.2 hold. If $\sqrt{n}\lambda_n=o(1)$ and $p_n\log n/\sqrt{n}=o(1)$, then with probability tending to 1 the adaptive lasso estimator $\hat{\beta}^{\mbox{\tiny AL}}$ has the following properties: i) $\hat{\beta}^{\mbox{\tiny AL}}_2=0$; ii) and for any nonzero $p_{n,1}\times1$ constant vector $u_n$ with $\|u_n\|=1$, $\sqrt{n}u_n^\top(\hat{\beta}^{\mbox{\tiny AL}}_1-\beta_{0,1})/[2(u_n^\top A_{11}^{-1}V_{11}Z_{11}^{-1}u_n)^{-1/2}]$ converges in distribution to $N(0,1)$.
\end{thm}

We give out the proof of Theorem \ref{thm2} in the Appendix A.2. By choosing a suitable $u_n$, it can be seen from Theorem \ref{thm2} that the estimator of each individual component of $\beta_{0,1}$ is asymptotically normal. The results can be used to make inference on individual coefficients. The estimation of the variance is discussed in Section 3.4.

\subsection{The selection of tuning parameter and variance estimation} 

Typical methods of selecting tuning parameter include information criteria methods such as AIC, BIC, and data-driven procedures such as $K$-fold cross-validation. Because of the computational issue, we consider information criteria approaches here. It is well known that BIC usually leads to the consistency of selection for fixed dimension scenarios. However, as we allow the dimension of the parameter to diverge, the standard BIC does not guarantee the selection consistency. Thus, we consider the following modified BIC:
\begin{eqnarray*}
	\mbox{BIC}(\lambda)=l_n(\hat{\beta}^{\mbox{\tiny AL}}_\lambda)+\frac{\log n}{n}C_n\mbox{df}_\lambda,
\end{eqnarray*}
where $\hat{\beta}^{\mbox{\tiny AL}}_\lambda$ represents the adaptive lasso estimate obtained at the tuning parameter with level $\lambda$, $C_n$ is a sequence going to infinity with $(\log n/n)C_n\to0$, and $\mbox{df}_\lambda$ stands for the size of the non-zero component of the estimate at level $\lambda$. We find that $C_n=\log\log p_n$ works well in this paper. The tuning parameter we choose is given by 
\begin{eqnarray*}
	\hat{\lambda}=\mbox{argmin}_\lambda\mbox{BIC}(\lambda).
\end{eqnarray*}
The proposed modified BIC has the consistency of selection. Let ${\cal M}^\ast$ and ${\cal M}_{\hat\lambda}$ be the true model and the selected model, respectively. We have the following result on selection consistency.

\begin{thm}\label{thm3}
	Assume that the conditions {\it C1} to {\it C6} listed in the Appendix A.2 hold. If $\sqrt{n}\lambda_n=o(1)$, $\lambda_nn/p_n\to\infty$, $p_n\log n/\sqrt{n}=o(1)$, and $\sqrt{n/\{C_np_n\log n\}}\times\liminf_{n\to\infty}\{\min_{j\in{\cal M}^\ast}|\beta_{0,j}|\}\to\infty$, then $\mathsf{P}({\cal M}_{\hat\lambda}={\cal M}^\ast)\to1$ as $n\to\infty$.
\end{thm}

Finally, we also need to give out the variance estimation of the proposed estimator for the nonzero component of $\beta_0$. Note that the asymptotic variance-covariance matrix of $\hat{\beta}^{\mbox{\tiny AL}}_1$ involves the density function of $\tvar$ and is complicated to estimate directly. We turn to use a random weighting method to obtain the variance estimate. To be specific, we defined the perturbed penalized estimator, denoted by $\hat{\beta}^\ast$, to be the minimizer of the following perturbed version of the objective function $l^{\mbox{\tiny AL}}_n(\beta)$:
\begin{eqnarray*}
	l_n^\ast(\beta)=\sum_{i=1}^n\sum_{j=1}^n(W_i+W_j)h_{ij}(\beta)+\hat{\lambda}\sum_{j=1}^{p_n}\hat{w}_j|\beta_j|,
\end{eqnarray*}
where $W_i$, $i=1,\ldots,n$ are i.i.d. non-negative and bounded random variables with mean 0.5 and variance 1, independent of the observed data. Let $\hat{\beta}^\ast_1$ be component of $\hat{\beta}^\ast$ corresponding to $\hat{\beta}^{\mbox{\tiny AL}}_1$, i.e., the non-zero part of $\hat{\beta}^{\mbox{\tiny AL}}$. It can be shown that under suitable conditions, given the observed data, the conditional asymptotic distribution of $\hat{\beta}^\ast_1$ is the same as the asymptotic distribution of $\hat{\beta}^{\mbox{\tiny AL}}_1$. Thus, by generating repeatedly the sequence of $W_i$, $i=1,\ldots,n$, we can obtain a large number of replications of $\hat{\beta}^\ast_1$ and then approximate the variance of $\hat{\beta}^{\mbox{\tiny AL}}_1$.

\section{Numerical results}

\subsection{Simulation studies}

We conduct some simulation studies to illustrate the finite sample performance of the proposed approach. We set $p_n=[7\tilde{n}^{1/5}]$, where $\tilde{n}$ is the size of the full sample, as defined in Section 2. The size of the non-zero components is set to be $p_{n,1}=[p_n/3]$. We consider two scenarios with $\tilde{n}=300$ and $\tilde{n}=500$, respectively. In the first scenario, $p_n=21$ and $p_{n,1}=7$. We set $\beta_{1,0}=(3.12, 2.20,-0.86,0.92,-2.49,1.95,-1.32)^\top$. For the covariates $\tx=(\tx_1,\ldots,\tx_{21})$, we independently generate $\tx_1$ from Binomial(0.25) (the Binomial distribution with success probability 0.25), $\tx_2$ from Binomial(0.8), $\tx_8$ from Uniform(0,2) (the uniform distribution from 0 to 2), $\tx_9$ from Binomial(0.5), and $\tx_{10}$ from Uniform(-2,0). All the rest covariats are generated from the standard normal distribution, with the correlation between $\tx_i$ and $\tx_j$ being $0.3^{|i-j|}$. In the second scenario, $p_n=24$ and $p_{n,1}=8$. We set $\beta_{1,0}=(3.12, 2.20,-0.86,0.92,-2.49,1.95,-1.32,-2.13)^\top$. We generate $\tx_1$ from Binomial(0.25), $\tx_2$ from Binomial(0.8), $\tx_9$ from Uniform(0,2), $\tx_{10}$ from Binomial(0.5), $\tx_{11}$ from Uniform(-2,0), and the rest from the standard normal distribution, with the correlation between $\tx_i$ and $\tx_j$ being $0.3^{|i-j|}$. For the error distribution in the linear model, we consider the standard normal ($N(0,1)$) distribution and the extreme value distribution with location parameter 0 and scale parameter 1 ($EV(0,1)$). The left truncation variable $\tl$ is generated from Uniform($a$,$b^\top X$) and the right one $\tr=\tl+c$, where $a$, $b$, and $c$ are constants to yield truncation percentage of 30\% and 40\%. The percentages of left and right truncation are almost the same. 

Under each scenario, 1000 replications are carried out. For variable selection, we define several criteria to assess the performance. The first one is the model error (ME), being defined as ME$=(\tilde{\beta}-\beta_0)^\top\mathsf{E}(X_iX_i^\top)(\tilde{\beta}-\beta_0)$, where $\tilde{\beta}$ is an estimator of $\beta_0$ derived from a specific approach. The median and median absolute deviation (MAD) of the 1000 MEs are recorded to evaluate the prediction performance of different procedures. The second one is the average number of total estimated zero coefficients (TN), that is, the average number of zero estimates obtained in the 1000 replications, including correctly estimated zero number (CN) and incorrectly estimated zero number (IN). For CN, the closer it is to $p_n-p_{n,1}$ the better a procedure performs, while for IN, the closer it is to 0 the better the procedure performs. The third one is the ratio of the correctly identified models (RCM), that is, the proportion of the replications that can identify the true model correctly among the 1000 replications. The closer it is to 100\%, the better a procedure performs. We consider three procedures, including the proposed selection procedure (Proposed), the variable selection procedure with adaptive lasso penalty without taking truncation into consideration (Naive), and the oracle procedure where the correct subset of covariates is used to fit the model (Oracle). The results are summarized in the following tables.

\begin{center}
	[Insert Table 1 here]
\end{center}

\begin{center}
	[Insert Table 2 here]
\end{center}

From the tables, we see that the proposed method outperforms the naive approach significantly in terms of the ME. This is expectable since the naive approach yields biased estimators for the regression coefficients. For the TN and RCM, the proposed method also has better results than the naive approach, while the extend of the improvement is not very large. The TN and RCM are criteria for variable selection performance. Although selecting important variables seems to be less affected by the double truncation, correcting the bias induced by double truncation still helps in increasing the precision of the variable selection procedure.

\subsection{A real data example}

We analyze an astronomical data from the 46420 quasar catalog produced by the Sloan Digital Sky Survey (SDSS) team. In this data, the main purpose is to predict the red shift of a quasar by using some related features. Thus, the linear model is used and the dependent variable is the red shift of the quasars. Compared with the complete SDSS catalog, the data set we use here omits some technical columns and therefore contains 23 covariates. Some more details about the data set can be found in {\it https://astrostatistics.psu.edu/datasets}.

It might not be that straightforward to see why there exists double truncation in the red shift data. This mainly comes from the nature of the observation feature of the astronomical telescope. Because of the aperture of the astronomical telescope, the red shift of a quasar one can observe is limited from the above. For the current data, we consider the maximum value of the red shift (5.4135) as the right truncation bound of the dependent variable. That is, the red shift larger this value is right truncated. For the low red shift, the measurement might be unreliable. The values should be omitted if the value of the red shift is lower its estimated error. The estimated error is then treated as the left truncation bound of the red shift observation. Therefore, the red shift value of the quasar is considered to be doubly truncated. Among the 23 covariates, some are the measurement errors of the brightness which are determined by the SDSS team from knowledge of the observing conditions, detector background, and other technical considerations. These covariates are not included in the model. We use 13 main covariates (shown in Table 3) to fit the linear model. Due to the computation capacity limitation, we do not use the full data to fit the model. A random sample with sample size 929 is drawn to illustrate the proposed method. The sampled data is split into the training part and testing part according to the proportion of 7:3. We use the proposed variable selection procedure on the training set to get a list of "important" covariates and then used the selected model to do prediction on the testing set. The results of the variable selection are summarized in Table 4. 

\begin{center}
	[Insert Table 3 here]
\end{center}

\begin{center}
	[Insert Table 4 here]
\end{center}

The proposed variable selection procedure gives out 5 "important" covariates with non-zero coefficients. We use the selected covariates to do the prediction on the testing set. In Figure 1, we draw the line plot to compare the predicted values and the true response values. We see that the predicted values (shown by the dotted line) are quite close to the true values (shown by the solid line). The selected covariates give out reasonable predictions. 

\begin{center}
	[Insert Figure 1 here]
\end{center}

\section{Concluding remarks}

We combine the Mann-Whitney-type loss function and the adaptive LASSO penalty to develop a simultaneous estimation and variable selection procedure for double truncated linear models, which allows the number of regression parameters to grow to infinity with the sample size. To overcome the non-convexity of the objective function, an iterative algorithm is designed. In each iteration, the objective function to be optimized can be transformed into a LAD problem. The oracle property of the proposed variable selection procedure is proved. We use the modified BIC to select the tuning parameter and adopt random weighting approach for variance estimation. Simulation studies show the reasonable finite sample performances of the proposed approach.

One of the main contributions of the current approach is to allow the dimension of the covariates to go to infinity with the sample size. However, by using the penalty based method, the growing rate of the covariates dimension can not be very large. For the ultra-high dimension situations, some screening approach is needed first. This is one of the interested directions that can be explored in future study. 

\appendix

\section{Appendix}

\subsection{More details on the algorithm}

Here we give out some more details on the algorithm discussed in Section 3.2. Define $\check{Y}=Y_i-Y_j$ and $\check{X}=X_i-X_j$ for $(i,j)\in{\cal S}_k$. Let $m$ be an appropriate index of ${\cal S}_k$. Thus $\sum_{i,j\in{\cal S}_k}|e_i(\beta)-e_j(\beta)|=\sum_{m=1}^N|\check{Y}_m-\beta^\top\check{X}_m|$. Further define the matrices
\begin{equation*}
\begin{matrix}
\check{\mathbf{Y}}=\begin{pmatrix}
Y_{i_1}-Y_{j_1} \\ 
\vdots\\
Y_{i_N}-Y_{j_N}
\end{pmatrix} ,&   
~~~\check{\mathbf X}(\beta)= 
\begin{pmatrix}
\beta^\top(X_{i_1}-X_{j_1}) \\ 
\vdots\\
\beta^\top(X_{i_N}-X_{j_N}) 
\end{pmatrix}
\end{matrix}
\end{equation*}
and the augmented matrices
\begin{equation*}
\begin{matrix}
\check{\mathbf Y}_{\mbox{\tiny aug}}=\begin{pmatrix}
\check{\mathbf Y}/n(n-1) \\ 
\mathbf{0}_{p_n}
\end{pmatrix} ,&   
~~~\check{\mathbf X}_{\tiny{\mbox{aug}}}(\beta,\lambda_n) = 
\begin{pmatrix}
\check{\mathbf X}(\beta)/n(n-1)\\
\mathbf{diag}(\lambda_n\hat{w})
\end{pmatrix}
\end{matrix},
\end{equation*}
where $\mathbf{0}_p$ is a $p_n$-vector of zeros and $\mathbf{diag}(\lambda_n\hat{w})$ is a $p_n\times p_n$ diagonal matrix with the elements $(\lambda_n\hat{w}_1,\cdots,\lambda_n\hat{w}_{p_n})$.
Based on these notations, to minimize the objective function in (\ref{steoLP}) can be formulated to minimize $\|\check{\mathbf Y}_{\mbox{\tiny aug}}- \check{\mathbf X}_{\tiny{\mbox{aug}}}(\beta,\lambda_n)\|_1$ with respect to $\beta$. This is a linear progamming problem and can be solved quite efficiently through standard unpenalized LAD procedure, such as \textbf{quantreg} package in \textbf{R}. The specific estimation procedure can be summarized as follows.

\noindent(i) Get an initial estimate value, for instance, the estimate value obtained by ignoring the double truncation and the regularization.

\noindent(ii) Iteratively solve the unpenalized optimization problem until convergence to get $\hat{\beta}$.

\noindent(iii) Use $\hat{\beta}$ to construct the objective function $ l_n^{\mbox{\tiny AL}}(\beta)$ and the corresponding modified one $ l_n^{\mbox{\tiny AL}}(\beta,b)$.

\noindent(iv) Use $\hat{\beta}^{(0)}$ as the initial value. Iteratively solve the optimization problem $\min_\beta\|\check{\mathbf Y}_{\mbox{\tiny aug}}- \check{\mathbf X}_{\tiny{\mbox{aug}}}(\beta, \lambda_n)\|_1$ by using linear programming until convergence to get $ \hat{\beta}^{\mbox{\tiny AL}}$.

\subsection{Proof of the large sample properties}

To obtain the large sample properties presented in Section 3.3, some regularity conditions are needed. Let $Z=(Y,R,L,X)$ be the population of the observed data. Let $\cal N$ be an $o(1)$ neighborhood of $\beta_0$ and $S$ be the sample space of $Z$. We use the notation $h(Z_i,Z_j;\beta)$ to represent $h_{ij}(\beta)$, where $Z_i=(Y_i,R_i,L_i,X_i)$, $i=1,\ldots,n$. Define $\tau(z,\beta)=\mathsf{E}[h(z,Z;\beta)]$. We use $\nabla_k$ to represent the $k$-th order derivative of a function of $\beta$ with respect to $\beta$. We assume the following conditions hold.

\noindent{\it C1}. The density function $f$ is bounded and has a bounded and continuous derivative.

\noindent{\it C2}. The parameter space $\Theta$ is compact and the true parameter value $\beta_0$ is an interior point of $\Theta$.

\noindent{\it C3}. Each component of the covariate vector has a bounded second moment. 

\noindent{\it C4}. For all $z\in S$, $\nabla_2\tau(z;\beta)$ exist on $\mathcal{N}$ and continuous at $\beta_{0}$.

\noindent{\it C5}. Let $I(z) =\sup_{\alpha \in S_p}\alpha^{\top}\nabla_2\tau(z; \beta_{0})\alpha$ with $S_p=\{\alpha\in R^{p_n}:\|\alpha\|=1 \} $.  $\mathsf{E}[I^2(Z)]<\infty $.

\noindent{\it C6}. There exist positive constants $C_1, C_2, C_3$, and $C_4 $ such that for all  $n$,
\begin{equation}
0<C_1<\lambda_{min}(A)\leqslant\lambda_{max}(A)<C_2<\infty, 0<C_3<\lambda_{min}(V)\leqslant\lambda_{max}(V)<C_4<\infty,
\end{equation}
where $\lambda_{min}(\cdot)$ and $\lambda_{max}(\cdot) $ are the largest and smallest eigenvalue of a matrix, respectively.

{\it C1} and {\it C2} are mild conditions. {\it C2} is used to prove the consistency of $\hat{\beta}$, which is crucial to guarantee the consistency of the proposed estimator $\hat{\beta}^{\tiny\mbox{AL}}$. {\it C3} is sufficient to meet the condition 5 in Wang et al. (2019) so that some results on high dimensional Hoeffding's decomposition can used to obtain the quadratic approximation of $l_n(\beta)-l_n(\beta_0)$. {\it C4} is for the Taylor's expansion of $\tau(z;\beta)$ at $\beta_0$. {\it C5} is used to bound the empirical term in the Hoeffding's decomposition of $l_n(\beta)-l_n(\beta_0)$. {\it C6} is used to confine the Hessian matrix of the quadratic approximation of the objective function and ensure that the covariance matrix of the score function is positively definite with uniformly bounded eigenvalues for all $n$. This provides justification for the component-wise asymptotic normality of $\hat{\beta}^{\tiny\mbox{AL}}$. Similar conditions can be found in some related literature, e.g., Peng and Fan (2004), Cai et al. (2005), Cho and Qu (2013), and Ni et al. (2016).   

We then give out the proof of the proposition and the theorems.

\noindent{\it Proof of Proposition 1}: For a U-process of order 2 with the kernel function $h\in{\cal H}$, where $\mathcal{H} =\{h(\cdot;\beta): \beta\in\Theta\}$ is a measurable class of symmetric functions on $ S^2\equiv S\otimes S$ with a non-negative envelope. Denote a set as $\mathbf{h}_\omega(\cdot;\beta)=\{h(Z_1(\omega),Z_2(\omega);\beta),\cdots, h(Z_{n-1}(\omega),Z_n(\omega);\beta)\} $ with length $n(n-1)$ and a class $\mathcal{H}_\omega =\{\mathbf{h}_\omega(\cdot;\beta):\beta\in\Theta\}$. According to the Example 1 in Wang et al. (2019) and Lemma 4.4 in Pollard (1990), we have that the pseudo-dimension of $\mathcal{H}_\omega$ is at most $100p_n$ for every $\omega$. If $p_n\log n/\sqrt{n}=o(1)$, it is sufficient to guarantee the condition on the pseudo-dimension in Proposition 1 of Wang et al. (2019). Hence, we have the following quadratic approximation
\begin{equation*}
l_n(\beta)-l_n(\beta_{0})=\frac{1}{2}(\beta-\beta_{0})^{\top}A(\beta-\beta_{0})+(\beta-\beta_{0})^{\top}\hat{U}_n+o_p(\|\beta-\beta_{0}\|^2)+o_p\left(\frac{\sqrt{p_n}\|\beta-\beta_{0}\|}{\sqrt{n}}\right),
\end{equation*}
where $\hat{U}_n=2\sum_{i=1}^{n}\xi_{ij|i}/n$ with $\xi_{ij|i}=\mathsf{E}[\xi_{ij}(\beta_0)|Z_i]$. By constraining $p_n$, the pseudo-dimension is at the rate of $o(\sqrt{n}/\log(n))$. Therefore, all the conditions required listed in Theorem 3 in Wang et al. (2019) are satisfied. Then the consistency of $\hat{\beta}$ can be obtained.\qed

To prove Theorem 1, a lemma is needed.

\begin{lem}
	Assume that the conditions {\it C1} to {\it C6} hold. Then $\forall b\in R^{p_n}$ with $\| b\|_2=1$, $\sqrt{n}(b^{\top}Vb)^{-1/2}b^{\top}\hat{U}_n/2$ converges in distribution to $N(0,1)$.
\end{lem}

\noindent{\it Proof}: It is not difficult to see that
\begin{equation*}
\frac{\sqrt{n}}{2}(b^{\top}Vb)^{-1/2} b^{\top}\hat{U}_n=\frac{1}{\sqrt{n}}\sum_{i=1}^{n}(b^{\top}Vb)^{-1/2}b^{\top} \xi_{ij|i}=:I_1.
\end{equation*}
Note that $\mathsf{E}(\xi_{ij|i})=E(\xi(Z_i,Z_j,\beta_0))=0$ and $Var(I_1)=1$. By {\it C3} and {\it C4}, $b^{\top}Vb$ is bounded. By the Cauchy-Schwarz inequality, we have that 
\begin{eqnarray*}
	\frac{1}{n^2}\sum_{i=1}^{n}\mathsf{E}\left|(b^{\top}Vb)^{-1/2}\cdot b^{\top} \xi_{ij|i}\right|^4&=&O(1)n^{-2}\sum_{i=1}^{n}\mathsf{E}\left|b^{\top} \xi_{ij|i}\right|^4
	\leqslant O(1)n^{-2}\sum_{i=1}^{n}\| \xi_{ij|i} \|_2^4\\
	&=&O_p(n^{-2}\cdot n \cdot p_n^{2})=O_p(\frac{p_n^2}{n})=o_p(1).
\end{eqnarray*}
Thus, the Lyapunov condition holds for $I_1$. By Lyapunov central limit theorem, $I_1$ converges in distribution to $N(0,1)$.\qed

\noindent{\it Proof of Theorem 1}:  
Let $\alpha_n=(p_n/n)^{1/2}$. It can be shown that for any $\epsilon>0$ and any constant vector $u_n$ with $\|u_n\|=C $, there exists a large enough $C$ such that 
\begin{equation*}
\mathsf{P}\left\{\inf_{\|u_n\|=C}l^{\mbox{\tiny AL}}_n(\beta_{0}+\alpha_n u_n)>l^{\mbox{\tiny AL}}_n(\beta_{0})\right\}\geqslant1-\epsilon.
\end{equation*}
This implies the existence of a local minimizer $\hat{\beta}^{\mbox{\tiny AL}}$ such that $\|\hat{\beta}^{\mbox{\tiny AL}}-\beta_{0} \|=O_p(\alpha_n)$. Then we have that 
\begin{eqnarray*}
	l^{\mbox{\tiny AL}}_n(\beta_{0}+\alpha_n u_n)-l^{\mbox{\tiny AL}}_n(\beta_0)&=& l_n( \beta_{0} + \alpha_n u_n)-l_n(\beta_0)+\sum_{j=1}^{p_n}\{P_{\lambda_nj}(|\beta_{0j} + \alpha_n u_{nj}|) -P_{\lambda_nj}(|\beta_{0j}|) \}\\
	&\geqslant&\{l_n(\beta_{0}+\alpha_n u_n)-l_n(\beta_0)\}+\sum_{j=1}^{p_{n,1}}\{P_{\lambda_nj}(|\beta_{0j} + \alpha_n u_{nj}|) -P_{\lambda_nj}(|\beta_{0j}|) \}\\
	&=:&U_1-U_2.
\end{eqnarray*}
We write that
\begin{equation*}
U_1=\alpha_n u_n^\top\hat{U}_n+\frac{1}{2}\alpha^2_n u_n^\top Au_n+(\|u_n\|^2+\|u_n\|)o_p(\alpha^2)=:U_{11}+U_{12}+U_{13},
\end{equation*}
where
\begin{equation*}
|U_{11}|=|\alpha_nu_n^\top\hat{U}_n|\leqslant\alpha_n\|u_n\|\|\hat{U}_n\|=\alpha_n\|u_n\|O_p\left(\sqrt{\frac{p_n}{n}}\right)=\|u_n\|O_p(\alpha_n^2)
\end{equation*}
and
\begin{equation*}
|U_{12}|=|\frac{1}{2}\alpha^2_n u_n^\top Au_n|\geqslant\frac{1}{2}\alpha^2_n\|u_n\|^2 \lambda_{min}(A)>\alpha^2_n\|u_n\|^2 C_1/2.
\end{equation*}
Therefore, for large enough $\|u_n\|$, $|U_{12}|$ dominates $|U_{11}|$ and $|U_{13}|$ in $U_1$. Meanwhile, 
\begin{eqnarray*}
	|U_2|&=&\left|\sum_{j=1}^{p_{n,1}}\frac{\lambda_n}{|\hat{\beta}_j|}|\beta_{0j}+\alpha_nu_{nj}|-\frac{\lambda_n}{|\hat{\beta}_j|}|\beta_{0j}|\right|\leqslant\alpha_n\lambda_n\left|\sum_{j=1}^{p_{n,1}}\frac{|u_{nj}|}{|\hat{\beta}_j|}\right|\\
	&=&\alpha_n\lambda_n\sqrt{p_{n,1}}O_p(1)\|u_n\|\|u_n\| o_p(\alpha^2_n). 
\end{eqnarray*}
For large enough $ C=\|u_n\|$, $|U_{12}|$ dominates $|U_{2}|$. Since $A$ is positive definite, then $U_{12}>0$. Consequently, $l^{\mbox{\tiny AL}}_n(\beta_{0}+\alpha_nu_n)-l^{\mbox{\tiny AL}}_n(\beta_{0})>0$ with probability tending to 1 as $n\to \infty $ for large enough $C$.\qed

To prove Theorem 2, another lemma is needed.

\begin{lem}\label{lem2}
	Assume that the conditions {\it C1} to {\it C6} hold. If $\sqrt{n}\lambda_{n}=o(1)$, $\lambda_{n}n/p_n\to\infty$, and $p_n\log n/\sqrt{n}=o(1)$, then with probability tending to 1, for any given $\beta_{1}$ satisfying $\|\beta_{1}-\beta_{0,1}\|=O(p_n^{1/2}n^{-1/2}) $ and any constant $C$, $l^{\mbox{\tiny AL}}_n\{(\beta_{1}^\top,0^\top)^\top\} =\min_{\|\beta_{2}\|\le Cp_n^{1/2}n^{-1/2}}l^{\mbox{\tiny AL}}_n\{(\beta_{1}^\top,\beta_{2}^\top)^\top\}$.
\end{lem}

\noindent{\it Proof}: It is sufficient to show that with probability tending to 1, for any $\beta_{1}$ satisfying $\|\beta_{1}-\beta_{0,1}\|=O(p_n^{1/2}n^{-1/2})$ and $\|\beta_{2}\|\leqslant Cp_n^{1/2}n^{-1/2}$, $\partial l^{\mbox{\tiny AL}}_n(\beta)/\partial\beta_{j}$, $\beta_{j}$ have the same signs for $j=(p_{n,1}+1),\cdots,p_n $. We have that 
\begin{eqnarray*} 
	\frac{\partial l^{\mbox{\tiny AL}}_n(\beta_n)}{\partial \beta_{j}}&=&\hat{U}_n(\beta_{0})_j+\sum_{k=1}^{p_n}A_{jk}(\beta_{k}-\beta_{0k})+o_p\left(\|\beta_n-\beta_{0}\|+\sqrt{\frac{p_n}{n}}\right)+P^\prime_{\lambda_nj}(|\beta_{j}|)sign(\beta_{j})\\
	&=:&V_1+V_2+V_3+V_4.
\end{eqnarray*} 
It can be shown that
\begin{equation*}
V_1=O_p(\frac{1}{\sqrt{n}})=o_p\left(\sqrt{\frac{p_n}{n}}\right),
\end{equation*}
\begin{equation*}
|V_2|=\left|\sum_{k=1}^{p_n}A_{jk}(\beta_{k}-\beta_{0k})\right|\leqslant\|\beta-\beta_{0}\|\left(\sum_{k=1}^{p_n}A_{jk}^2\right)^{\frac{1}{2}}=O_p\left(\sqrt{\frac{p_n}{n}}\right)O(1)=O_p\left(\sqrt{\frac{p_n}{n}}\right),
\end{equation*}
and $|V_3|=o_p(\sqrt{p_n/n})$. Thus, $V_1+V2+V_3=O_p(\sqrt{p_n/n})$. Following this, we have that
\begin{eqnarray*}
	\frac{\partial l^{AL}_n(\beta_n)}{\partial \beta_{j}}&=&P^\prime_{\lambda_nj}(|\beta_{j}|)sign(\beta_{j}) + O_p\left(\sqrt{\frac{p_n}{n}}\right)=\frac{\lambda_n}{|\hat{\beta}_{nj}|}sign(\beta_{j})+O_p\left(\sqrt{\frac{p_n}{n}}\right)\\
	&=&\frac{\lambda_n}{|\hat{\beta}_{nj}|} \left\{ sign(\beta_{j})+ O_p\left(\frac{p_n}{n\lambda_n}\right)\right\}.
\end{eqnarray*}
Since $\lambda_nn/p_n\to\infty$, $sign(\beta_{j})$ dominates $O_p(p_n/(n\lambda_n))$. Therefore, $\partial l^{\mbox{\tiny AL}}_n(\beta_n)/\partial \beta_{j} $ and $ \beta_{j} $ have the same signs for $ j=(p_{n,1}+1),\cdots,p_n $  with probability tending to 1 as $ n\to\infty$.\qed

\noindent{\it Proof of Theorem 2}: The part i) follows directly from Lemma \ref{lem2}. To prove ii), we first denote that  $B_n=\{P^\prime_{\lambda_{n1}}(|\beta_{01}|)sign(\beta_{01}),\cdots,P^\prime_{\lambda_{np_{n,1}}}(|\beta_{0k}|)sign(\beta_{0p_{n,1}})\}^\top$. It is easy to see that $B_n$ converges to 0 as $n\to\infty$. Since $\hat{\beta}^{\mbox{\tiny AL}}$ is the minimizer of $l^{\mbox{\tiny AL}}_n(\beta)$, $\partial l^{\mbox{\tiny AL}}_n(\hat{\beta}^{\mbox{\tiny AL}})/\partial\beta_{1}=0$. Combining $\hat{\beta}^{\mbox{\tiny AL}}_{2}=0$, we get that 
\begin{equation}\label{oracle}
\hat{U}_{1}+A_{11}(\hat{\beta}^{\mbox{\tiny AL}}_{1}-\beta_{0,1})+o_p(\|\beta_n-\beta_{0}\|+\sqrt{\frac{p_n}{n}})+B_n=0,
\end{equation}
where  $\hat{U}_{1}$ consists of the first $p_{n,1}$ components of $\hat{U}_{n}$. Rearrange (\ref{oracle}) to get
\begin{equation*}
\sqrt{n}(\hat{\beta}^{\mbox{\tiny AL}}_{1}-\beta_{0,1})=-\sqrt{n}A_{11}^{-1} \hat{U}_{1}-\sqrt{n}A_{11}^{-1} B_n+o_p(\sqrt{p_n}).
\end{equation*}	
By the condition that $\sqrt{n}\lambda_{n}=o(1)$, we have that $\sqrt{n}A_{11}^{-1}B_n\to0$. By {\it C1} and {\it C3}, it can be seen that  $\mathsf{E}|\nabla_{1j}\tau(Z_1'\beta_{0})|^2<\infty$ for $j=1,\cdots,p_n$. Thus, $\|A_{11}^{-1/2}\hat{U}_{1}\|=O_p(\sqrt{p_n/n})$ and
\begin{equation*}\label{per}
\sqrt{n}(\hat{\beta}^{\mbox{\tiny AL}}_{1}-\beta_{0,1})=\sqrt{n}A_{11}^{-1} \hat{U}_{1}(1+o_p(1)).
\end{equation*}
Since for any nonzero $p_{n,1}\times1 $ constant vector $ u_n $ with $\|u_n\|=1$,
$\sqrt{n}u_n^\top A_{11}^{-1} \hat{U}_{1}/[2\left[u_n^\top A_{11}^{-1}V_{11}A_{11}^{-1}u_n\right]^{-1/2}]$ converges in distribution to $N(0,1)$. Thus, by Slutsky's theorem, we conclude that $\sqrt{n}u_n^\top (\hat{\beta}^{\mbox{\tiny AL}}_{1}-\beta_{0,1})/[2\left[u_n^\top A_{11}^{-1}V_{11}A_{11}^{-1}u_n\right]^{-1/2}]$ converges in distribution to $N(0,1)$.\qed

\medskip

\noindent{\it Proof of Theorem 3}: To state the proof more clearly, some more notation are needed. The underlying true model is denoted by $\mathcal{M}^\ast= \{j: \beta_{0j}\ne0, j=1,\ldots,p_n\}$. If we use $\hat{\beta}_{\lambda}$ to represent the estimator at the tuning parameter value $\lambda$, then the corresponding model is denoted by  $\mathcal{M}_\lambda=\{j: \hat{\beta}_{\lambda,j}\ne0, j=1,\ldots,p_n\} $. For any model $\mathcal{M}=\{j_1,\ldots,j_{m}\}$ with size $m$, the corresponding estimated coefficients are $\hat{\beta}_{\mathcal{M}}=(\beta_{j_1},\ldots,\beta_{j_m}) $. Note that $\hat{\beta}_{\mathcal{M}}$ and $\hat{\beta}_{\lambda}$ are two different terms. We need to distinguish them carefully. Next, we divide the $R^p$ space into the following three disjoint regions: i) $\Omega_{\minus}=\{\lambda>0: \mathcal{M}_\lambda\not\supset\mathcal{M}^\ast\}$, ii) $\Omega_{0}=\{\lambda>0 : \mathcal{M}_\lambda=\mathcal{M}^\ast\}$, and iii) $\Omega_{+}=\{\lambda>0: \mathcal{M}_\lambda\supset\mathcal{M}^\ast, \mathcal{M}_\lambda\ne\mathcal{M}^\ast\}$. This division divides the range of $\lambda$ into 3 categories, corresponding to the under-fitted model, the correctly fitted model and the over-fitted model. For the under-fitted model, there exists $j\in\mathcal{M}^\ast$ and $j\notin\mathcal{M}_\lambda$. The over-fitted model refers to the model such that for all $j\in\mathcal{M}^\ast$, $j\in\mathcal{M}_\lambda$, but there exists $k\in \mathcal{M}_\lambda$ and $k\notin\mathcal{M}^\ast$. We prove the theorem by considering the two parts: the under-fitted model and the over-fitted model.

For the under-fitted model part, when $\lambda\in\Omega_{\minus}$, the length of the vector $\hat{\beta}^{\mbox{\tiny AL}}_S$ is inconsistent with $\beta_0$. In order to make $\hat{\beta}^{\mbox{\tiny AL}}_S$ and $\beta_0$ comparable, let us slightly modify the definition of $\hat{\beta}_S$ here. In this part, we define for any model $S$,  
\begin{equation*}
\hat{\beta}_{S} =\mbox{argmin}_{\beta\in R_{k_n}:\beta_j=0,\forall i\notin S}l_n (\beta).
\end{equation*}
By definition, 
\begin{eqnarray*}
	\inf_{\lambda \in \Omega_{\minus}}\mbox{BIC}_\lambda&=&\inf_{\lambda \in \Omega_{\minus}}l_n(\hat{\beta}_\lambda) +\frac{\log n}{n}C_n \cdot \mbox{df}_\lambda\geqslant\inf_{\lambda \in \Omega_{\minus}} l_n(\hat{\beta}_\lambda)\\
	&=&\inf_{\lambda\in\Omega_{\minus}}l_n(\beta_0) +\frac{1}{2}(\hat{\beta}_\lambda^\ast-\beta_{0})^{\top}A(\hat{\beta}_\lambda^\ast-\beta_{0})\geqslant l_n(\beta_0)+\frac{1}{2}\lambda_{min}(A)\inf_{\lambda \in \Omega_{\minus}}\|\hat{\beta}_\lambda^\ast-\beta_{0}\|^2\\
	&\geqslant&l_n(\beta_0)+\frac{1}{2}\lambda_{min}(A)\inf_{S\not\supset \mathcal{M}^\ast}\|\hat{\beta}_S-\beta_{0} \|^2\geqslant l_n(\beta_0)+\frac{1}{2}\lambda_{min}(A)\min_{j \in \mathcal{M}_*}(\beta_{0,j}^2).
\end{eqnarray*}
By {\it C6}, we can obtain that 
\begin{equation}\label{A2}
\inf_{\lambda \in \Omega_{\minus}}\mbox{BIC}_\lambda-l_n(\beta_0)\geqslant\frac{1}{2}C_1\min_{j \in\mathcal{M}^\ast}(\beta_{0,j}^2).
\end{equation} 
By the condition $\sqrt{n/\{C_n p_n\log n\}}\times\liminf_{n\to\infty}\{\min_{j\in S_T}|\beta_{0,j}| \}\to\infty$, the term on the right side of (\ref{A2}) is $O(C_np_n\log n/n)$, which is $O(\log\log p_n\times p_n\log n/n) $ as we choose $C_n=\log \log p_n $ here. Similarly, we derive that
\begin{eqnarray*}
	\mbox{BIC}_{\hat{\lambda}}&=&l_n(\hat{\beta}) +\frac{\log n}{n}C_n\cdot \mbox{df}\leqslant l_n(\hat{\beta})+\frac{p_n\log n\log\log p_n}{n}\\
	&=&l_n(\beta_0) +\frac{1}{2}(\hat{\beta}-\beta_{0})^{\top}A(\hat{\beta}-\beta_{0}) +\frac{p_n\log n\log\log p_n}{n}\\
	&\leqslant&l_n(\beta_0) +\frac{1}{2}\lambda_{max}(A) \|\hat{\beta}-\beta_{0}\|^2+\frac{p_n\log n\log\log p_n}{n}\\
	&=&l_n(\beta_0) + O_p\left(\dfrac{p_n}{n}\right)+\frac{p_n\log n \log \log p_n}{n}\\
	&=&l_n(\beta_0) +o_p\left(\frac{p_n\log n\log\log p_n}{n}\right)+\frac{p_n\log n\log\log p_n}{n}
\end{eqnarray*}
and
\begin{equation*}
\mbox{BIC}_{\hat{\lambda}}-l_n(\beta_0)=O_p\left(\frac{p_n\log n\log\log p_n}{n}\right).
\end{equation*}
Combining the two results, we can obtain that
\begin{equation}\label{under}
\mathsf{P}\left(\inf_{\lambda \in \Omega_{\minus}}\mbox{BIC}_\lambda>\mbox{BIC}_{\hat{\lambda}}\right)\to1.
\end{equation}

For the over-fitted model part, for any $\lambda\in\Omega_+$,
\begin{eqnarray*}
	\inf_{\lambda \in \Omega_+}\left(\mbox{BIC}_\lambda-\mbox{BIC}_{\hat{\lambda}}\right)&=&\inf_{\lambda\in \Omega_+}l_n(\hat{\beta}_\lambda)-l_n(\hat{\beta})+(\mbox{df}_\lambda- \mbox{df}_{\hat{\lambda}})\frac{\log n}{n}C_n\\
	&=&\inf_{\lambda\in\Omega_+}\frac{1}{2}(\hat{\beta}_\lambda-\beta_{0})^{\top}A(\hat{\beta}_\lambda-\beta_{0})-\frac{1}{2}(\hat{\beta}-\beta_{0})^{\top}A(\hat{\beta}-\beta_{0})\\
	& &+(\mbox{df}_\lambda-\mbox{df}_{\hat{\lambda}})\frac{\log n}{n}C_n\\
	&\geqslant&-\frac{1}{2} \lambda_{max}(A) \|\hat{\beta}-\beta_{0} \|^2+(\mbox{df}_\lambda-\mbox{df}_{\hat{\lambda}})\frac{\log n}{n}C_n\\
	&\geqslant&-\frac{1}{2}C_2\|\hat{\beta}-\beta_{0} \|^2+\frac{\log n}{n}C_n.
\end{eqnarray*}
The last line holds since for any $\lambda\in\Omega_+$, it must satisfy $\mbox{df}_\lambda-\mbox{df}_{\hat{\lambda}}\geqslant1$. Furthermore, as $\|\hat{\beta}^{\mbox{\tiny AL}}-\beta_0\|^2 =O_p(p_n/n)$, so
\begin{equation*} 
\inf_{\lambda \in \Omega_+}\left(\mbox{BIC}_\lambda-\mbox{BIC}_{\hat{\lambda}}\right)\geqslant(1+ o_p(1))\frac{p_n\log n\log\log p_n}{n}>0.
\end{equation*}
This means that
\begin{equation}\label{over}
\mathsf{P}\left(\inf_{\lambda\in\Omega_{+}}\mbox{BIC}_\lambda>\mbox{BIC}_{\hat{\lambda}}\right)\to 1.
\end{equation}
Combining (\ref{under}) and ( \ref{over}) we can conclude that:
\begin{equation*}
\mathsf{P}(\mathcal{M}_{\hat{\lambda}}=\mathcal{M}^\ast)\to1
\end{equation*}
as $n\to\infty$.\qed

\bigskip

\noindent {\bf Acknowledgment}

The research of Ming Zheng was supported by the National Natural Science Foundation of China Grants (11771095). The research of Wen Yu was supported by the National Natural Science Foundation of China Grants (12071088). 

\medskip

\section*{References}

\begin{enumerate}[{[}1{]}]
\item Bhattacharya, P. K., Chernoff, H., and Yang, S. S. (1983), Nonparametric estimation of the slope of a truncated regression, {\it The Annals of Statistics}, 11, 505-514.
\item Bilker, W., and Wang, M.-C. (1996), Generalized Wilcoxon statistics in semiparametric truncation models, {\it Biometrics}, 52, 10-20.
\item Cai, J., Fan, J., Li, r., and Zhou, H. (2005). Variable selection for multivariate failure time data. {\it Biometrika}, 92, 303?16.
\item Cho, H. and Qu, A. (2013), Model selection for correlated data with diverging number of parameters. {\it Statistica Sinica}, 23, 901?27.
\item Efron, B., and Petrosian, V. (1999), Nonparametric methods for doubly truncated data, {\it Journal of the American Statistical Association}, 94, 824-834.
\item Greene, W. H. (2012), {\it Econometric Analysis (7th Ed.)}, Prentice Hall, Upper Saddle River, NJ.
\item Keiding, N., and Gill, R. D. (1990), Random truncation models and Markov processes, {\it The Annals of Statistics}, 18, 582-602.
\item Kim, J.P., Lu, W., Sit, T. and Ying, Z. (2013), A unified approach to semiparametric transformation models under generalized biased sampling schemes, {\it Journal of the American Statistical Association}, {108}, 217-227.
\item Lai, T. L., and Ying, Z. (1991a), Estimating a distribution function with truncated and censored data, {\it The Annals of Statistics}, 19, 417-442.
\item Lai, T. L., and Ying, Z. (1991b), Rank regression methods for left-truncated and right-censored data, {\it The Annals of Statistics}, 19, 531-556.
\item Liu, H., Ning, J., Qin, J. and Shen, Y. (2016), Semiparametric maximum likelihood inference for truncated or biased-sampling data'', {\it Statistica Sinica}, 26, 1087-1115.
\item Lynden-Bell, D. (1971), A method of allowing for known observational selection in small samples applied to 3CR quasars, {\it Monthly Notices of the Royal Astronomical Society}, 155, 95-118.
\item Moreira, C. and Una-Alvarez, J. (2012), Kernel density estimation with doubly truncated data, {\it Electronic Journal of Statistics}, 6, 501-521.
\item Moreira, C. and Una-Alvarez, J. (2016), Nonparametric regression with doubly truncated data, {\it Computational Statistics and Data Analysis}, 93, 294-307.
\item Ni, A., Cai, J., and Zeng, D. (2016), Variable selection for case-cohort studies with failure time outcome, {\it Biometrika}, 103, 547-562.
\item Peng, H. and Fan, J. (2004). Nonconcave penalized likelihood with a diverging number of parameters, {\it The Annals of Statistics}, 32, 928-961.
\item Pollard, D. (1990), {it Empirical Processes: Theory and Applications}, IMS,Hayward, CA.
\item Shen, P.-S., (2010), Nonparametric analysis of doubly truncated data, {\it Annals of the Institute of Statistical Mathematics}, 62, 835-853.
\item Shen, P.-S. (2013a), A class of rank-based tests for doubly-truncated data, {\it TEST}, 22, 83-102.
\item Shen, P.-S. (2013b), Regression analysis of interval censored data and doubly truncated data. {\it Computational Statistics}, 28, 581-596.
\item Tsai, W.-Y. (1990), Testing the independence of truncation time and failure time, {\it Biometrika}, 77, 167-177.
\item Tsui, K.-L., Jewell, N. P., and Wu, C. F. J. (1988), A nonparametric approach to the truncated regression problem, {\it Journal of the American Statistical Association}, 83, 785-792.
\item Turnbull, B. W. (1976), The empirical distribution function with arbitrarily grouped, censored and truncated data, {\it Journal of the Royal Statistical Society}, Ser. B, 38, 290--295.
\item Wang, M.-C., Jewell, N. P., and Tsai, W.-Y. (1986), Asymptotic properties of the product limit estimate under random truncation, {\it The Annals of Statistics}, 14, 1597-1605.
\item Wang, Z., Lin, Y., Liu, W., and Shao, Q. (2019), U-processes with increasing dimensional parameters, working paper.
\item Ying, Z., Yu, W., Zhao, Z., and Zheng, M. (2020), Regression analysis of doubly truncated data,  {\it Journal of the American Statistical Association}, 115, 810-821. 
\item Zou, H. (2006), The adaptive lasso and its oracle properties.  {\it Journal of the American Statistical Association}, 101, 1418-1429.
\end{enumerate}

\clearpage

\begin{table}[h]   
\centering   
{\caption{Simulation results for variable selection when the random error follows the standard normal distribution.}}
\label{1}
\medskip
\begin{tabular}{*{8}{c}}
\toprule[1.0pt]	
\multirow{2}*{$\tilde{n}$}&\multirow{2}*{Truncation}&\multirow{2}*{Method}&\multicolumn{2}{c}{ME}&\multicolumn{2}{c}{TN}&\multirow{2}*{RCM}\\
\cmidrule(lr){4-5}\cmidrule(lr){6-7}&percentage& & Median & MAD & CN & IN &(\%)\\
\midrule
\multirow{6}*{300}	&	\multirow{3}*{0.3}	&	 Proposed 	&	0.220	&	0.137	&	13.921	&	0.000	&	92.836	\\
\cmidrule{3-8}															
&	 	&	 Naive 	&	0.487	&	0.242	&	13.908	&	0.000	&	91.364	\\
\cmidrule{3-8}															
&	 	&	 Oracle 	&	0.109	&	0.076	&	14.000	&	0.000	&	100.000\\	
\cmidrule{2-8}															
&	\multirow{3}*{0.4}	&	 Proposed 	&	0.385	&	0.232	&	13.862	&	0.000	&	88.679	\\
\cmidrule{3-8}															
&	 	&	 Naive 	&	1.048	&	0.423	&	13.821	&	0.000	&	84.608	\\
\cmidrule{3-8}															
&	 	&	 Oracle 	&	0.158	&	0.112	&	14.000	&	0.000	&	100.000	\\
\midrule															
\multirow{6}*{500}	&	\multirow{3}*{0.3}	&	 Proposed 	&	0.133	&	0.073	&	15.995	&	0.000	&	99.503	\\
\cmidrule{3-8}															
&	 	&	 Naive 	&	0.289	&	0.135	&	15.966	&	0.000	&	96.723	\\
\cmidrule{3-8}															
&	 	&	 Oracle 	&	0.062	&	0.043	&	16.000	&	0.000	&	100.000	\\
\cmidrule{2-8}															
&	\multirow{3}*{0.4}	&	 Proposed 	&	0.217	&	0.120	&	15.977	&	0.000	&	98.014	\\
\cmidrule{3-8}															
&	 	&	 Naive 	&	0.611	&	0.229	&	15.938	&	0.000	&	94.048	\\
\cmidrule{3-8}															
&	 	&	 Oracle 	&	0.084	&	0.057	&	16.000	&	0.000	&	100.000	\\
\bottomrule[1.0pt]															
\end{tabular}
\end{table}

\begin{table}[h]  
	\centering  
	\caption{Simulation results for variable selection when the random error follows the extreme minimum value distribution.}  
	\label{2}
	\medskip
	\begin{tabular}{*{8}{c}}
		\toprule[1.0pt]	
		\multirow{2}*{$ \tilde{n}$}&\multirow{2}*{Truncation}&\multirow{2}*{Method}&\multicolumn{2}{c}{ME}&\multicolumn{2}{c}{TN}&\multirow{2}*{RCM}\\
		\cmidrule(lr){4-5}\cmidrule(lr){6-7}&percentage& & Median & MAD & CN & IN &(\%)\\
		\midrule
		\multirow{6}*{300}	&	\multirow{3}*{0.3}	&	 Proposed 	&	0.226	&	0.162	&	13.85	&	0.00	&	86.905	\\
		\cmidrule{3-8}															
		&	 	&	 Naive 	&	0.780	&	0.353	&	13.82	&	0.00	&	83.730	\\
		\cmidrule{3-8}															
		&	 	&	 Oracle 	&	0.129	&	0.098	&	14.00	&	0.00	&	100.000	\\
		\cmidrule{2-8}															
		&	\multirow{3}*{0.4}	&	 Proposed 	&	0.380	&	0.293	&	13.81	&	0.00	&	83.765	\\
		\cmidrule{3-8}															
		&	 	&	 Naive 	&	1.232	&	0.565	&	13.72	&	0.00	&	75.896	\\
		\cmidrule{3-8}															
		&	 	&	 Oracle 	&	0.184	&	0.132	&	14.00	&	0.00	&	100.000	\\
		\midrule															
		\multirow{6}*{500}	&	\multirow{3}*{0.3}	&	 Proposed 	&	0.118	&	0.087	&	15.98	&	0.00	&	98.214	\\
		\cmidrule{3-8}															
		&	 	&	 Naive 	&	0.489	&	0.208	&	15.96	&	0.00	&	95.933	\\
		\cmidrule{3-8}															
		&	 	&	 Oracle 	&	0.075	&	0.048	&	16.00	&	0.00	&	100.000	\\
		\cmidrule{2-8}															
		&	\multirow{3}*{0.4}	&	 Proposed 	&	0.197	&	0.146	&	15.90	&	0.00	&	91.144	\\
		\cmidrule{3-8}															
		&	 	&	 Naive 	&	0.989	&	0.369	&	15.88	&	0.00	&	89.154	\\
		\cmidrule{3-8}															
		&	 	&	 Oracle 	&	0.107	&	0.072	&	16.00	&	0.00	&	100.000	\\
		\bottomrule[1.0pt]												
	\end{tabular}
\end{table}

\begin{table}[h] 
	\centering  
	\caption{Description of the 13 covariates to fit the linear model for the SDSS data}
	\label{SDSS data}
	\medskip
	\begin{tabular}{*{2}{l}}
		\toprule[1.0pt]
		\multicolumn{1}{c}{Name~~~~	} &Description~~~~ \\
		\midrule
		R.A.	&	Right Ascension 	\\	
		Dec.	& 	Declination 	\\
		u\_mag	& 	Brightness in the u (ultraviolet) band in magnitudes.	\\
		g\_mag	& 	Brightness in the g (green) band	\\
		r\_mag	&	Brightness in the r (red) band	\\
		i\_mag	&	Brightness in the i (more red) band	\\
		z\_mag	&	Brightness in the z (even more red) band	\\
		Radio	&	Brightness in the radio band	\\
		X-ray	&	Brightness in the X-ray band	\\
		J\_mag  &	Brightness in the near-infrared J band	\\
		H\_mag	&	Brightness in the near-infrared H band	\\
		K\_mag	&	Brightness in the near-infrared K band	\\
		M\_i	&   The absolute magnitude in the i band.	\\              										
		\bottomrule[1.0pt]									
	\end{tabular}
\end{table}

\clearpage

\begin{table}[h]  
	\centering  
	\begin{threeparttable}
		\caption{Estamtion and variabel selection results for the SDSS data}  
		\label{SDSS}
		\medskip
		\begin{tabular}{*{4}{c}}
			\toprule[1.0pt]		
			\multicolumn{2}{c}{	~~~~~~~Covariate~~~~	} &~~~~~EST~~~~ &~~~~SEE~~~~~\\
			\midrule
			1	&	R.A.	&	0	&	-	\\
			2	&	Dec.	&	0	&	-	\\
			3	&	u\_mag	&	0.340	&	0.024	\\
			4	&	g\_mag	&	-0.320	&	0.046	\\
			5	&	r\_mag	&	0	&	-	\\
			6	&	i\_mag	&	0.500	&	0.043	\\
			7	&	z\_mag	&	0	&	-	\\
			8	&	Radio	&	0	&	-	\\
			9	&	X.ray	&	0	&	-	\\
			10	&	J\_mag	&	0.017	&	0.007	\\
			11	&	H\_mag	&	0	&	-	\\
			12	&	K\_mag	&	0	&	-	\\
			13	&	M\_i	&	-0.803	&	0.017	\\												
			\bottomrule[1.0pt]
		\end{tabular}
		\small
		EST: estimation of coefficients, \\SEE: estimation of standard error
	\end{threeparttable}
\end{table}

\begin{figure}[h]
	\centering			
	\includegraphics[scale=0.45]{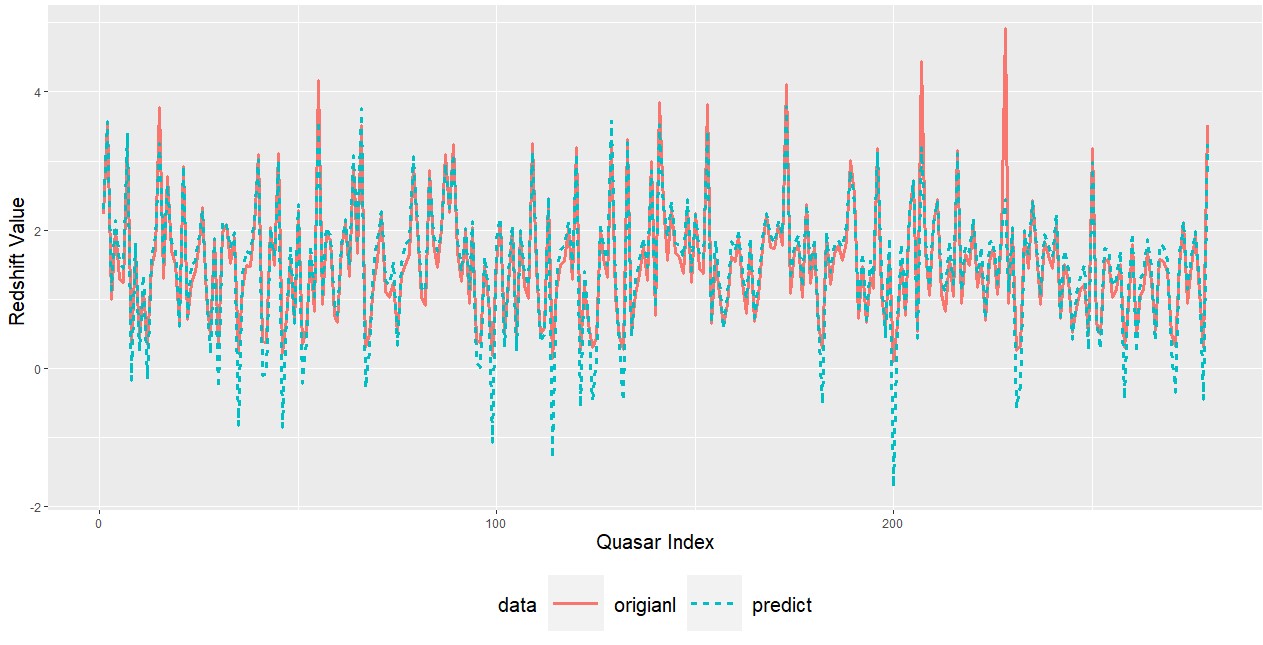}
	\caption{The predicted and original red shift values for the testing set}
	\label{fig:SDSS}
\end{figure}

\end{document}